\documentclass[%
 reprint,
 amsmath,amssymb,
 aps,
]{revtex4-2}

\usepackage{graphicx}
\usepackage{dcolumn}
\usepackage{bm}

\begin{document}

	\title{A universal signature in the melting of metallic nanoparticles}

	\author{Laia Delgado-Callico} 
	\affiliation{Department of Physics, King's College London, Strand, London WC2R 2LS, United Kingdom}
	\affiliation{These authors contributed equally to this work}

	\author{Kevin Rossi} 
	\affiliation{Laboratory of Computational Science and Modeling, Institute of materials, Ecole Polytechnique F\'{e}d\'{e}rale de Lausanne, Lausanne, 1015, CH.}
	\affiliation{Department of Physics, King's College London, Strand, London WC2R 2LS, United Kingdom}
	\affiliation{These authors contributed equally to this work}

	\author{Pascal Salzbrenner}%
	\affiliation{Department of Physics, King's College London, Strand, London WC2R 2LS, United Kingdom}
	\affiliation{University of Cambridge, Centre for Scientific Computing, Cavendish Laboratory, 19 J J Thomson Avenue, Cambridge CB3 0HE, United Kingdom}
	
	\author{Raphael Pinto-Miles}
	\affiliation{Department of Physics, King's College London, Strand, London WC2R 2LS, United Kingdom}
	
	\author{Francesca Baletto}
    \email{francesca.baletto@kcl.ac.uk}
	\affiliation{Department of Physics, King's College London, Strand, London WC2R 2LS, United Kingdom}
	
	\date{September 2020}%
	\revised{ 2020}%
	
	\begin{abstract}
Determining the melting transition temperature in metallic nanoparticles is a challenge from both the experimental and theoretical points of view. We reveal a universal criterion based on the distribution of the atomic pair-distances to characterise the melting behaviour of monometallic nanoparticles. 

We analyse the solid-liquid change of transition metals nanoparticles (Ni, Cu, Pd, Ag, Au and Pt) ranging from 146 to 976 atoms, considering several initial isomers for each nanoparticle and placing them in either vacuum or embedded in liquid environments simulated with an implicit force-field.
Regardless of the initial configuration, size and material, the peak from the second nearest-neighbours in the pair-distance distribution function disappears when the melting transition occurs. 
Therefore, melted and solid nanoparticles can be clearly distinguished because of a significant change in their pair distance distribution function, an experimentally measurable quantity.
We further show that the relative cross-entropy of the pair-distance distribution function between a ``cold” and a ``hot” reference structures correlates with the obtained caloric curves. 
The phase change temperature can then be easily estimated from the quasi-first order transition that the cross-entropy presents at the melting temperature.
	\end{abstract}
		\maketitle



\section{Introduction}

The novelty of nanotechnology stems from the unique physical and chemical properties of its building blocks, known as nanoparticles (NPs). Nanoparticles differ from both their atomic and bulk counterparts because of their large surface-to-volume ratio and of the coexistence of different isomers.
At first sight, there is an expectation that physical properties will show a smooth dependence with the NP size and shape.\cite{Guisbiers2019} 
Indeed, Pavlov’s suggestion \cite{Pavlov1903} of a linear depression of the melting point with the inverse of the NP diameter has been confirmed in several experiments.\cite{Schmidt2002} 
However, deviations from this general trend occur especially at small sizes. 
As a well-known example, the melting temperature of Na-NPs shows large fluctuations with NP-size,\cite{Kusche1999} and for Ga or Sn nanoparticles melting points can also exceed the bulk limit.\cite{Shvartsburg2000,Breaux2004,Steenbergen2016}


Therefore, Lord Kelvin’s question “Does the melting temperature of a small particle depend on its size?” \cite{Thomson1871} is still very actual. Several theoretical \cite{Tsallis1988,Hill1962,Hill2001}, numerical\cite{Li2008, Li2014, Li2013b, Qi2016, Calvo2015, Hou2017, Aguado2011,vasquez2015,Arslan2005} and experimental investigations   
\cite{Kusche1999, Shvartsburg2000, Breaux2004, Schmidt2002, Takagi1954, Buffat1976, Martin1996, Haberland1998, vanTeijlingen2020} have succeeded one another to capture the complexity inherent to the phase changes of a nanosized object. 
To answer it in full detail, we need to find measurable and computable physical quantities which probe when a solid-liquid (and \textit{viceversa}) phase change takes place.

The direct analysis of the caloric and heat capacity curves is a traditional numerical tool to infer when a phase change takes place. \cite{Li2008, Calvo2015, Pavan2015, Hou2017, Rossi2018a, Mottet2004} 
Molecular dynamics simulations can further provide insights into the appearance of kinetics factors during the phase transition. 
For example, it has been shown that the hysteresis of the melting- and the freezing- points of a metallic NP (MNP) is due to differences inherent to the mechanisms through which solid-liquid and solid-solid changes take place.\cite{Pavan2015, Hou2017, Rossi2018a} Furthermore, there is an evidence that the environment surrounding the MNP affects its thermodynamical properties, resulting, for example, in a pronounced smoothness of the caloric curve.\cite{Huerto-Cortes2013} 

Different experimental techniques such as electron diffraction,\cite{Takagi1954, Buffat1976} mass spectrometry, \cite{Martin1996} calorimetry \cite{Haberland1998} and optical spectroscopy \cite{Bosiger1987, Even1989, Buck1994, Ellert1995} relate the change of some physical quantities, i.e. optical/mass spectra or diffusion cross-sections, to a phase change in the nanoparticle.
Recently, van Teijlingen et al. \cite{vanTeijlingen2020} suggested that the solid-liquid phase change in spherical-like MNPs is characterized by a gradual expansion. Indeed, the authors reported transmission electron microscopy measurements which shown a surge in the averaged NP-diameter of at least 4.3\% at the melting.

Looking at differences in the structure of solid and liquid nanoparticles is also a robust method to detect phase changes. 
The comparison of the relative abundances of atoms displaying a specific and well defined local arrangement. \cite{Honeycutt1987a, Steinhardt1983, Baletto2005, Rossi2018a} is in fact an efficient way to discriminate phase changes with respect to solid-to-solid transitions.
Another popular method is based on probing the root mean bond fluctuations per each atom. 
A system is defined as melted if the average bond fluctuations are above a certain threshold, i.e. the Lindeman’s criterion.\cite{Lindemann}
Besides all the studies performed, the quest to identify a universal signature that marks a phase change of nanoparticles  and is common to both numerical simulations and experiments is still open and motivates our investigation.

Here, we show that melted and solid nanoparticles can be clearly distinguished because of a significant change in their pair distance distribution function (PDDF), an experimentally measurable quantity.\cite{Liao2018,Ingham2015}
We observe that the second peak of the PDDF disappears at the temperature where the energy has the biggest change, i.e., where the heat capacity vs temperature curve presents its maximum - the melting temperature.
We validated this observation by sampling the structural evolution of the noble and quasi-noble metallic nanoparticles with sizes in the 1-4 nm range, in vacuo and surrounded by an interacting environment, at various temperatures. 
We consistently observe the disappearance of the second peak in the PDDF in correspondence with the NP melting. Therefore, we consider this occurrence as a universal signature of phase change in metallic nanoparticles.
To quantitatively support this observation, we show the correlation between a nanoparticle caloric curve and the temperature-dependent relative cross-entropy of the pair distance distribution function between a ``cold” and a ``hot” reference structure. 
Moreover, we probe the influence of a strong interacting environment surrounding the nanoparticles confirming that our characterization method is as robust in a liquid-type environment as in vacuo.


\section{Simulation methods}

For a comprehensive structural characterization of transition metal nanoparticles undergoing phase changes we focus on a wide set of nanoparticles of diverse elements (Ni, Cu, Pd, Ag, Au and Pt), sizes and initial shapes (Figure S1). 
%
%
To simulate phase changes in metallic nanoparticles we 
concatenate canonical molecular dynamics runs where the temperature of the system is increased by a temperature $\Delta T$ every $\Delta \tau$ time. The ratio $\Delta T$ /$\Delta \tau$ tunes the heating rate, hence determining the kinetics of the solid-liquid phase change. \cite{Rossi2018a,Chen2020}
Our choice is a rate of 5~k/ns, with $\Delta T$ =25~K and $\Delta \tau$=5~ns. 
The Velocity-Verlet algorithm is used to evolve Newton’s Equation of motion.
The Andersen thermostat is applied to mimic the interaction of the system with a stochastic bath.

The initial and final temperatures used in the simulations depend on the particular system, and are adjusted such that the solid-liquid transition is clearly mapped while carrying out the simulations in an affordable time.

We model metal-metal interactions according to the Rosato-Guillope-Legrande \cite{Rosato1989} formulation 
 whose parametrizations for the metals here considered are reported elsewhere \cite{Baletto2002}.
To model the interaction of a metallic nanocluster with a non inert-environment, $E^{M - E}_{i}$, we adhere to the formalism introduced by Cortes-Huerto and coworkers.\cite{Huerto-Cortes2013} 
Here $E^{M - E}_{i}$ is calculated as a function of the number of absent bonds in surface atoms, with respect to their positioning in the bulk, $CN_{open}= CN_{bulk} - CN_{i}$, weighted by two free parameters, $\epsilon$ and $\rho$:
\begin{equation}
E^{M-E}_{i} = - \epsilon \mbox{~} CN_{open}^{\rho} \mbox{~~~,}
\label{Eq:hgn}
\end{equation}
where $CN_{bulk}$ equals to 12 for the case of FCC metals and $CN_{i}$ is calculated analytically \cite{Rossi2017a}
%

In this framework, the nature of the interaction between the atoms at the surface and the environment is encoded in the $\rho$ parameter. It takes a value of 1 for pairwise interaction, $<$~1 for covalent-like interactions, and $>$~1 for strongly interacting environments.
The $\epsilon$ parameter tunes the interaction strength instead.
Different $\rho$ and $\epsilon$ parameters set the ratio between the surface energies of low Miller index terminations, thus introducing a parameter to favour one architecture from another one. The simulations reported here employ  the following ($\rho$, $\epsilon$) pairs: (1, 0.02 eV), (1, 0.04 eV), (1, 0.08 eV), (1.5, 0.02 eV), (1.5, 0.04 eV) and (2, 0.02 eV).
While this model does not refer to any specific type of chemical conditions, we note that the metal-environment model for Au nanoparticles  employing parameters (1.5, 0.04 eV) and (2.0, 0.02 eV) was found to qualitatively match structural trends for the interaction of Au NP with a solution of Cetyltrimethylammonium bromide with silver ions (CTA$^+$/[AgBr$_2$]$^-$).\cite{Canbek2015}

\section{Energy and Characterization methods}

To analyse within a coherent framework the energetic trends found for nanoparticles of different size and composition we monitor their excess energy $\Delta$  \cite{Baletto2002}:
\begin{equation}
\Delta = (E_{\mathrm{tot}} - N E_ {\mathrm{coh}} ) / N^{2/3},
\end{equation}
with $E_{\mathrm{tot}}$ the total energy of the system, $N$ the number of atoms in the cluster, and $E_{\mathrm{coh}}$ the bulk cohesive energy.

To characterize the nanoparticles structural evolution when subject to heating up to melting and beyond, we monitor the distribution of the distances between each pair of atoms therein, i.e. their PDDF:
\begin{equation}
    g(d) = \frac{1}{(N)(N-1)} \sum_{i}^{N} \sum_{j \ne i}^{N} \delta (r_{ij} - d) ~~~~~~\mbox{,}
    \label{eq:PDDF}
\end{equation}
where r$_{ij}$ is the distance between two atoms $i$ and $j$ and $d$ is the distance at which the distribution is calculated.
The PDDF of a nanoparticle can be probed experimentally,\cite{Liao2018} and encodes rich information of the nanoparticle geometrical properties.\cite{Steenbergen2014,Zeni2018,rossi2017pccp,pavan2015jcp,amodeo2020}

Numerically, the resolution of the PDDF is dictated by the choice of the binning distance, which determines if two atoms are allocated into different bins (i.e. the width of the $\delta$ function in equation \ref{eq:PDDF}).
This choice needs to balance between a too high resolution - where each distance would be present a single time, hence the PDDF yields no useful information - and a too low one - where different neighbour shells are projected onto the same distance width, resulting in a too coarse description of the system.
Taking half of the distance between the first and second nearest neighbour as the largest distance bin width is a proper choice for distinguishing nearest neighbours peaks (Figure S2).

To ease the characterization of the PDDF-profile evolution at different temperatures, we estimate the Kullback-Leibler divergence, \cite{kullback1951} using the PDDF up to the second neighbour distance, between a reference temperature (from now on denominated as "cold", c) and a higher temperature (from now on labelled as "hot", h).
Commonly used in information theory, the Kullback-Leibler divergence (KL(h$\mid$c)) - also known as cross-entropy - establishes a quantification of the amount of information lost when a function c is used to approximate another function h. 
It takes values of 0 if h and c are equal and increases the more they differ. For a discrete distribution, it is calculated as:
\begin{equation}
	KL (h \mid  c) = \sum_i h(i) \log \dfrac{h(i)} {c(i)}.
	\label{eq:KL}
\end{equation}

During the dynamical evolution, we store instantaneous atomic positions, system excess energy, and temperature every 10~ps.
For a robust investigation we analyze results averaged from at least eight independent simulations.
We average the instantaneous excess energies of all the configurations in the same NVT ensemble and calculate the standard deviation from the average excess energy. 
Similarly, for each configuration we first calculate the corresponding PDDF, with distance occurrences grouped in bins of 0.1A. 
Finally, we estimate an average PDDF for each temperature by summing the instantaneous PDDF therein, normalized by the number of configurations.


\section{Results}

\begin{figure}[t!]
    \centering
    \includegraphics[width=8cm,height=18cm]{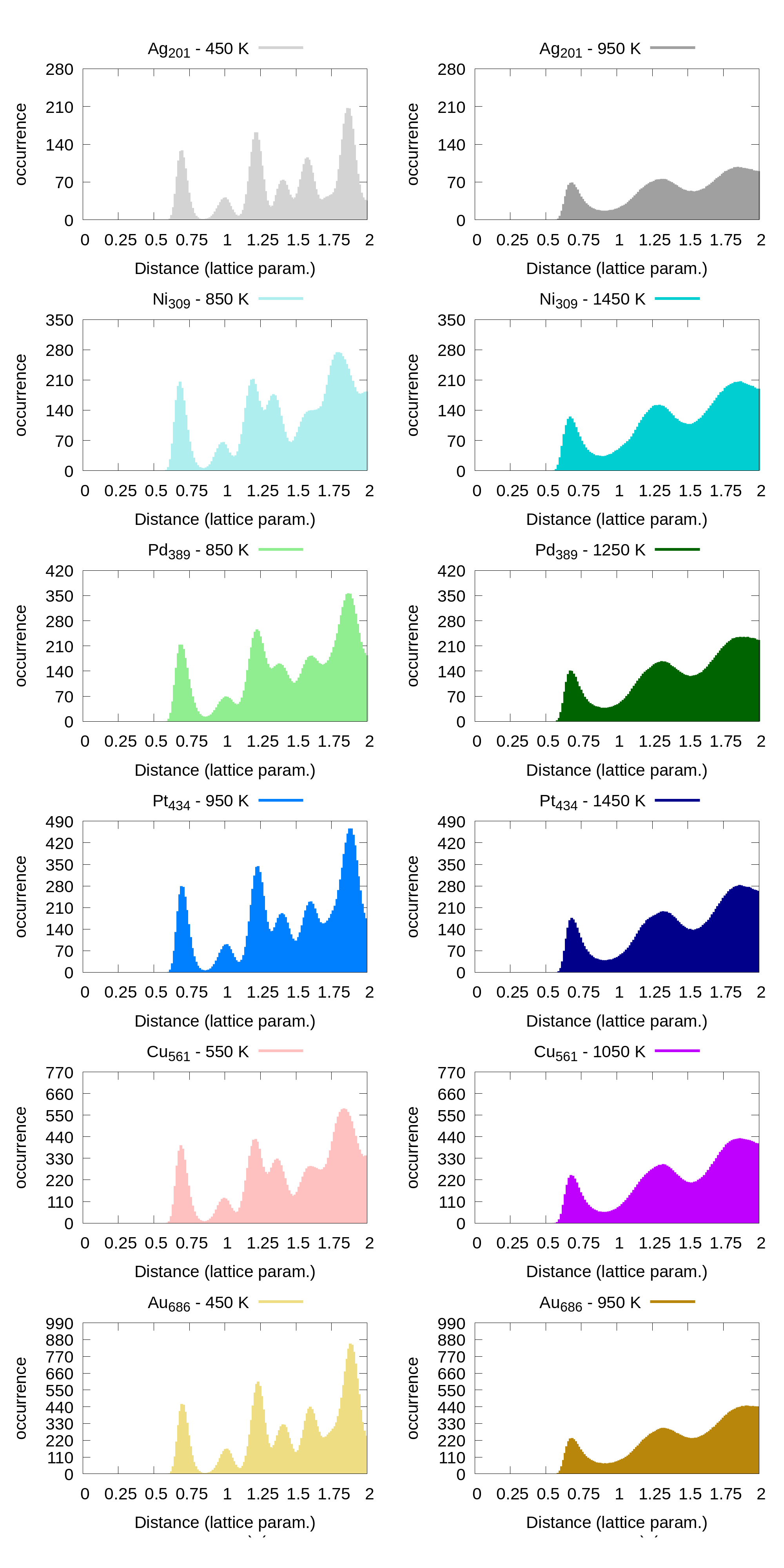}
    \caption{Average pair-distance distribution function of MNPs: for each temperature, we average over the instantaneous PDDF of the configuration sampled during the corresponding  time interval (5~ns).  Colours identify the different metals from top to bottom: Ag (silver), Ni (cyan), Pd (green), Pt (blue), Cu (magenta), Au (gold). Light colours on the left panel correspond to solid structure (``cold") while dark colours to the liquid phase (``hot") on the right panel.}
    \label{fig:PDDFs}
\end{figure}

A representative array of PDDFs for metallic nanoparticles of different sizes, shapes and compositions is reported in Figure \ref{fig:PDDFs}. The two panels refer to two different temperatures, namely ``cold" (left panel) and ``hot" (right panel). The reported PDDFs are averaged over at least eight independent simulations employing the same starting configuration but different initial velocities (yet coherent with the NVT ensemble of choice).
We note that the PDDF first peak falls close to, but is lesser than, the expected value of the nearest neighbours distance of a FCC $\sqrt{2}$/2 of the bulk lattice constant. 
At ``cold" temperatures, the second peak of the PDDF is instead close to the bulk lattice constant.
The other peaks of the PDDF depend on the specific NP geometry and size.
In the case of liquid droplets (``hot" temperature) depicted in the right panel of Figure \ref{fig:PDDFs}, we systematically observe the absence of a peak from the PDDF at the 2$^{nd}$ nearest-neighbours distance.
We also note that the maximum distance up to which the PDDF extends increases with temperature in systems that present an initial isotropic / spherical-like morphology (e.g. Icosahedra).
This is however not the case for systems which display an anisotropic (e.g. Octahedra) shape (Figure S3).

\begin{figure}[t!]
    \centering
     \includegraphics[width=8.cm]{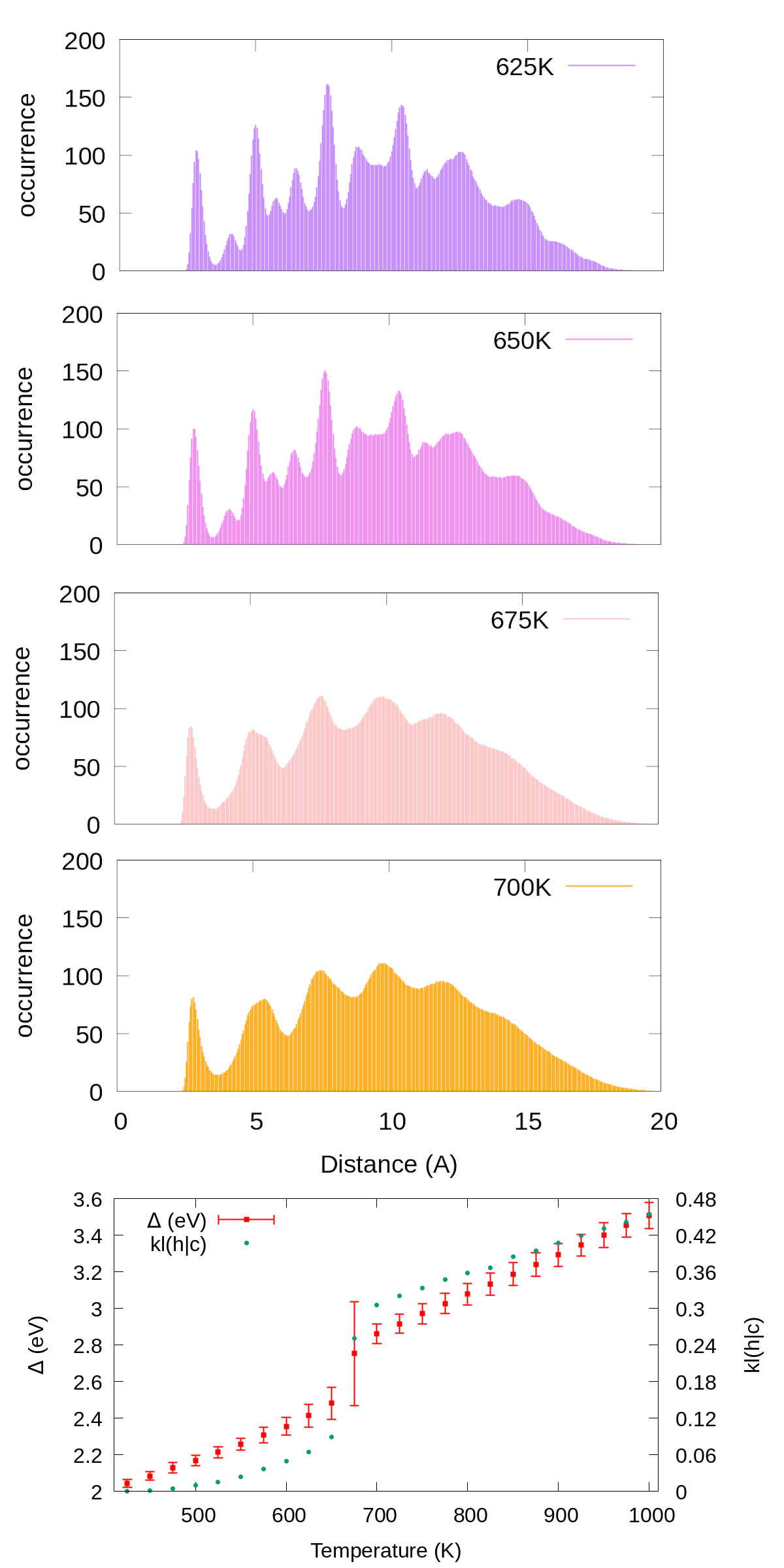}
    \caption{The upper panels show the average PDDF for Ag$_{201}$ taken at different temperatures. Above 675~K, the second peak of the PDDF is not at the bulk lattice value, with just a shoulder visible at $\sim$ 4.08A at 675~K. The bottom panel reports the caloric curve (red squares) and KL(h$\mid$c) (green circles) temperature dependence for a Ag nanoparticle of 201 atoms. A quasi first order transition occurs at 675~K.  }
    \label{fig:PDDFs_melt}
\end{figure}

Following a careful assessment of the temperature-evolution of the position of the PDDF peaks, we note that the PDDF second peak systematically and gradually diminishes in intensity at increasing temperatures (Figure S4-S26). 
Figure \ref{fig:PDDFs_melt} reports the paradigmatic example of the temperature dependent evolution of the PDDF in a Ag nanoparticle of 201 atoms, the peak at around 4.08\AA~ is clear at 625K, whilst it becomes a shoulder of the third neighbours' peak around 675K and vanishes above this temperature.
The latter corresponds to the melting temperature, $T_m$, as found from the caloric curve analysis, where we defined $T_m$ as the temperature at which the average excess energy, $\Delta$, presents the largest standard deviation.
We thus deem the absence of a peak at the second nearest-neighbours distances in the NP pair-distance distribution function to be a universal fingerprint for the solid-liquid transition.

To support this claim we contrasted the temperature-evolution of the position of the second peak of the PDDF against the caloric curve in all the systems under consideration.
The same trend is verified in all systems, independently of the initial size or shape of the nanoparticle. 
In turn, our finding establishes a fundamental relationship between the structure and the phase of a MNP.
Further, it enables the identification of the MNP phase from its pair distance distribution function, an accurate and well-defined quantity, accessible to both experimental and theoretical investigations.

Because of the temperature-driven changes in the PDDF of a MNP, its profile at high temperatures cannot be deduced from the one calculated at colder temperatures (and $\it{viceversa}$, but to a lesser extent).
In other words, the peaks of the "hot temperature" profile are wide and smooth, in contrast with the sharp peaks characteristic of the "cold temperature" distribution.
Conversely, when looking at the $\Delta$ vs Temperature and KL(h$\mid$c) vs Temperature curves (lower panel of Figure \ref{fig:PDDFs_melt}) we see that both show a quasi-discontinuity at the phase change temperature.
A positive correlation between the caloric and the KL(h$\mid$c) temperature dependent curves is not limited to the phase coexistence region, but also to the solid and liquid phases. 

We verify that any temperature choice below the melting transition does not affect the existence of a quasi-first order transition in the KL(h$\mid$c) vs temperature plot (e.g. Figure S27). 
We further confirm that the KL(h$\mid$c) discontinuity at the melting transition is independent of the choice of the distance bin width to calculate the nanoparticle PDDF, as long as this is smaller than 0.1 of the lattice parameter (Figure S28). 
Preliminary results on the KL robustness against the occurrence of structural rearrangements further show that the choice of limiting the cross-entropy calculation to the 2$^{nd}$ nearest neighbours is beneficial, with KL calculated for the full PDDF being instead less resilient (Figure S29).

\begin{figure}
    \centering
    \includegraphics[height=3.6cm]{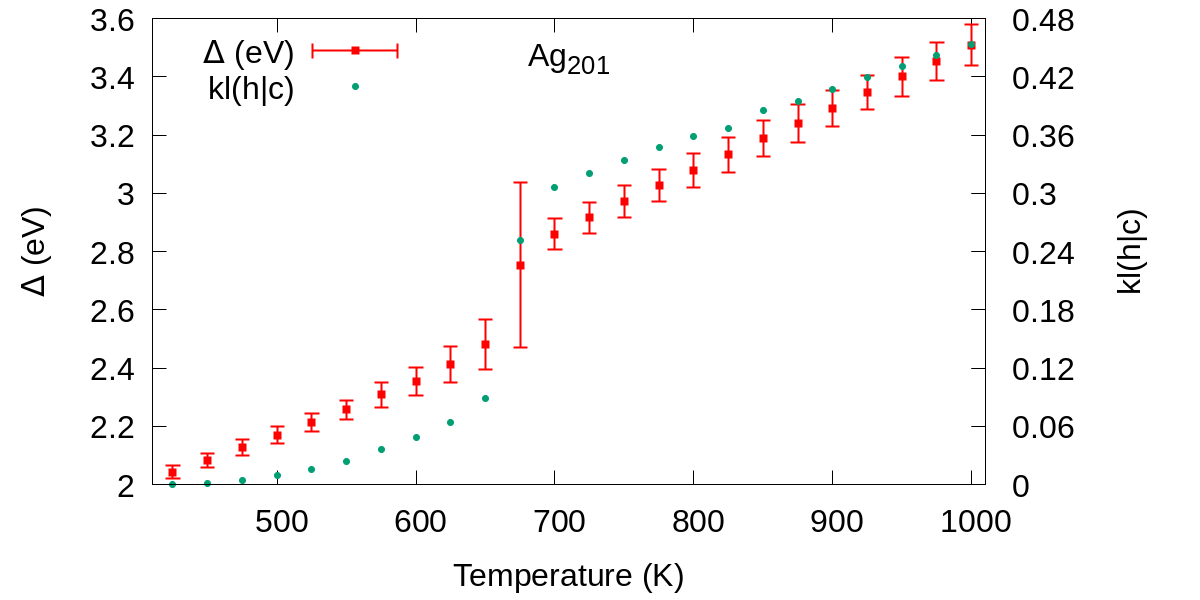}
    \includegraphics[height=3.6cm]{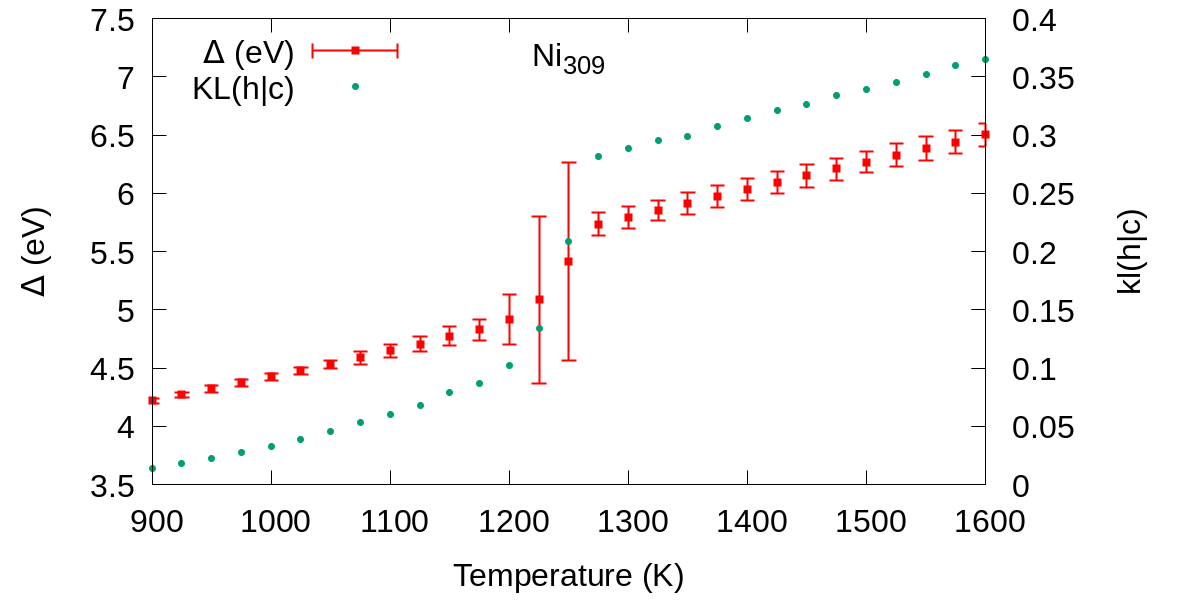}
    \includegraphics[height=3.6cm]{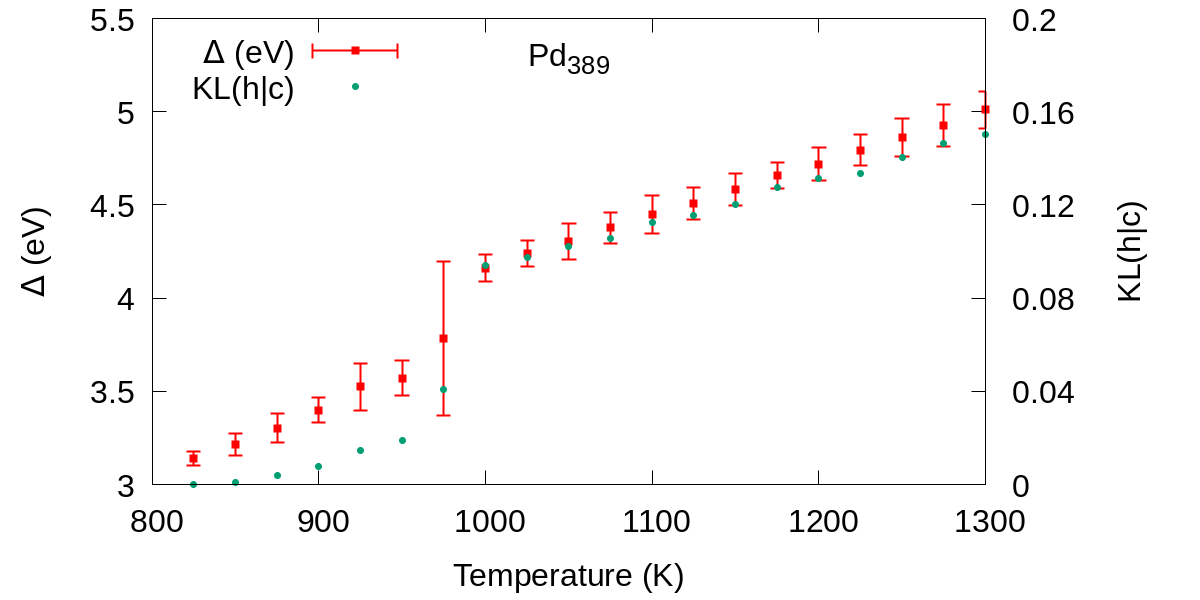}
    \includegraphics[height=3.6cm]{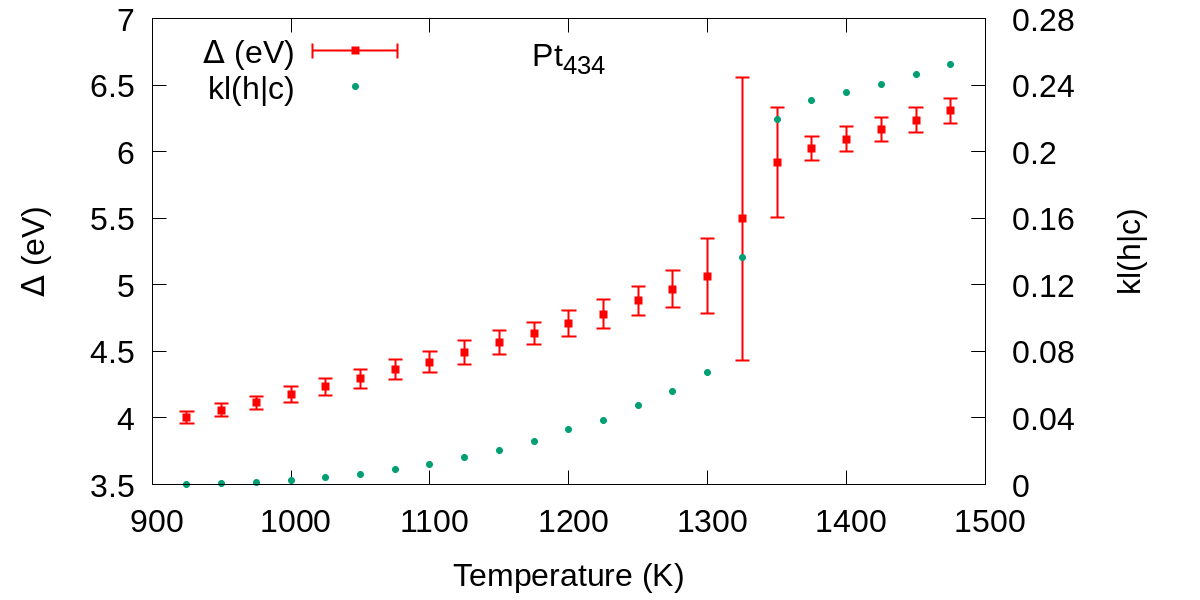}
    \includegraphics[height=3.6cm]{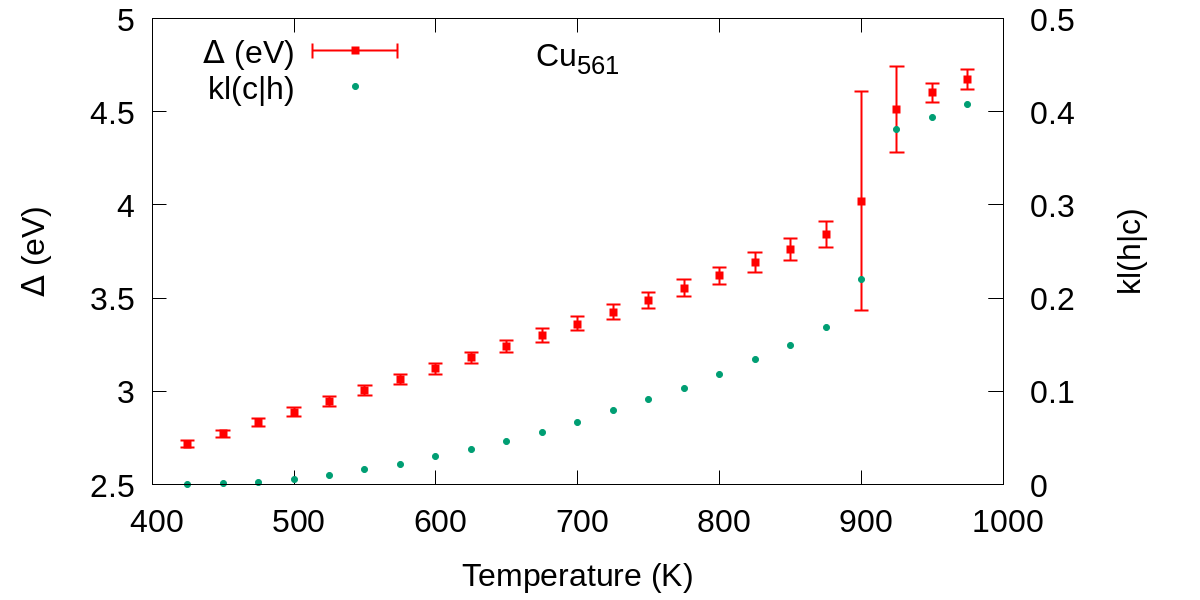}
    \includegraphics[height=3.6cm]{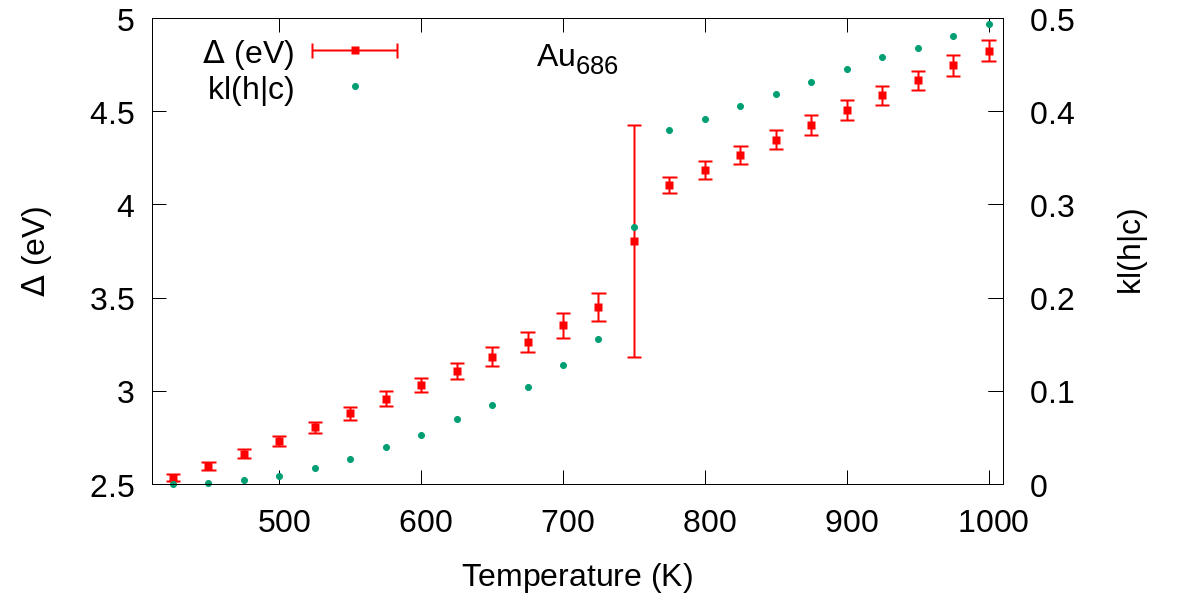}
    \caption{ Caloric curve (red squares) and KL(h$\mid$c) (green circles) dependence with respect to temperature in an array of systems of different size, composition, and initial structure. Both order parameters display a quasi first order transition at the melting.}
    \label{fig:kl_caloric_2}
\end{figure}

In Figure \ref{fig:kl_caloric_2} we show that the 1-to-1 correlation between the excess energy $\Delta$ and the KL(h$\mid$c) as a function of temperature holds true across a representative set of systems with different size, initial shape, and composition. 
The comparison of energy vs temperature and PDDF cross-entropy vs temperature for the systems here considered is provided in Figure S30 - S41. 
In each case, the KL PDDF analysis can be applied to probe solid-liquid phase transition in metallic nanoparticles.

\begin{figure}
    \centering
    \includegraphics[width=8cm]{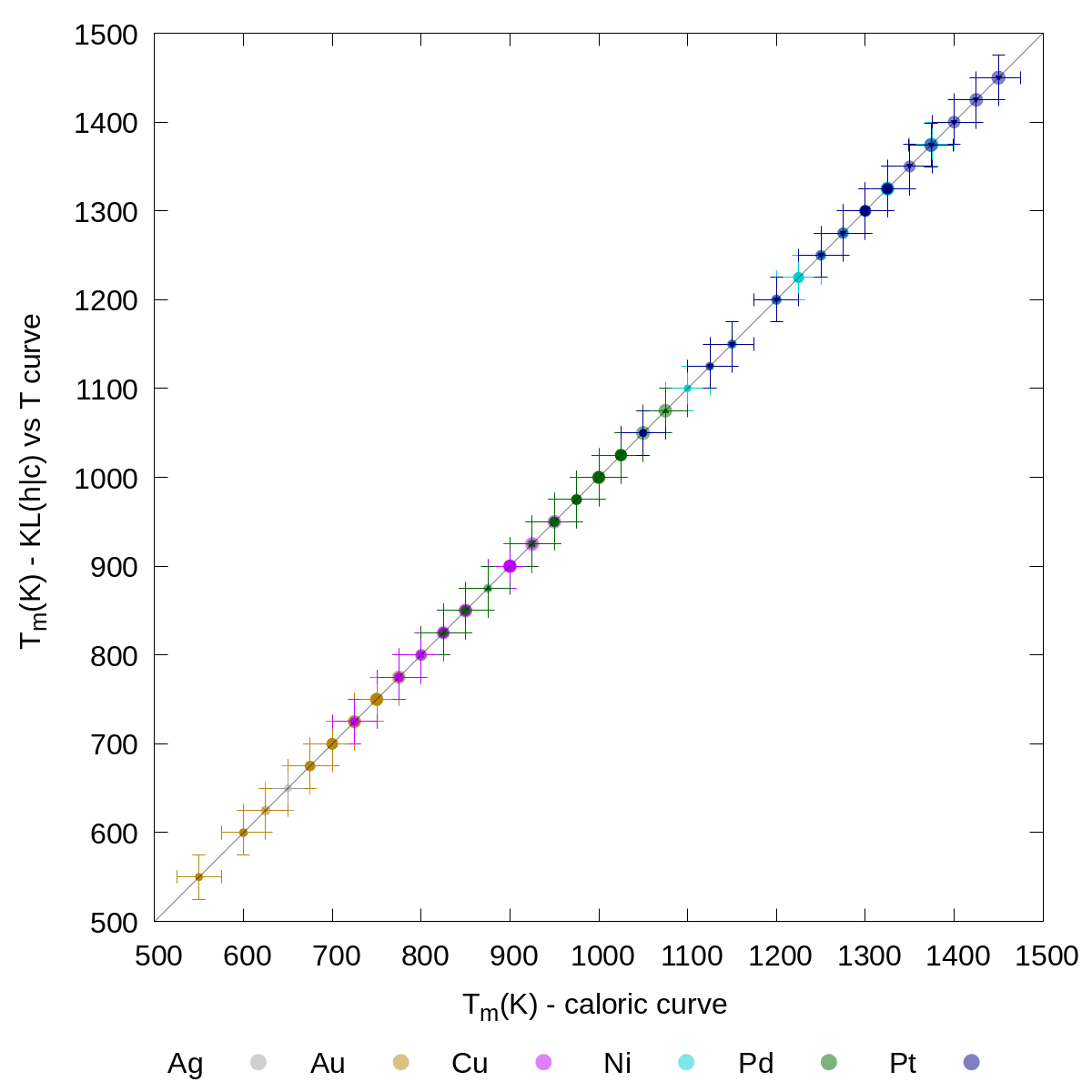}
    \caption{Parity plot between melting temperature, T$_{melting}$, estimated from the caloric curves (x axis) or the KL (h$\mid$c) vs temperature curves. Different systems compositions are colour coded according to the legend, while the point size is proportional to the cubic root of the nanoparticle size. The error bars are taken to be as $\Delta$T (25~K).}
    \label{fig:kl_caloric_3}
\end{figure}

In Figure \ref{fig:kl_caloric_3} we graphically highlight the 1-to-1 correspondence between melting temperature estimates according to the excess energy and the KL(h$\mid$c) evolution against temperature.
$T_m$ from the caloric curve is taken at the temperature where the largest standard deviation in the $\Delta$ curve occurs.
$T_m$ from the KL(h$\mid$c) temperature evolution is instead estimated as the temperature at which the KL(h$\mid$c) value is the most different from the ones reported for immediately preceding and subsequent temperatures.
For reference, we also report in Table S1 the T$_m$ found in each system under consideration according to both methods. 


\begin{figure}
    \centering
    \includegraphics[width=8cm]{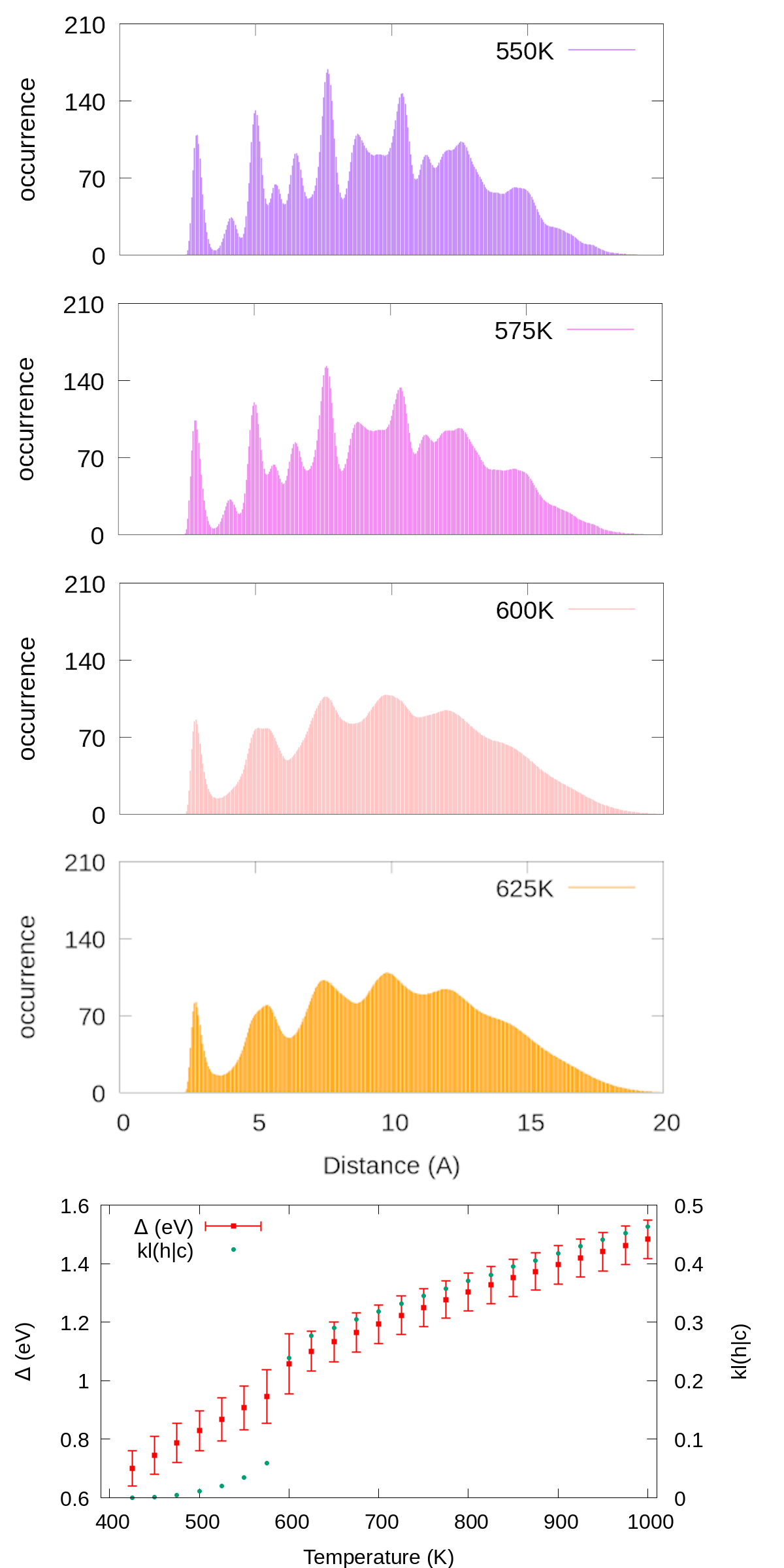}
    \caption{The upper panels show the average PDDF for Ag$_{201}$ embedded in a strongly interacting environment ($p$=1.5, $\epsilon$=0.04eV/atom) at different temperatures. Note the disappearance of the second peak of the PDDF for temperatures at 600~K. The lower panel shows Caloric curve (red squares) and KL(h$\mid$c) temperature dependence (green circles). A quasi first order transition in the KL(h$\mid$c) curve at 625~K signals the nanoparticle melting, yet this is less evident from the caloric curve.}
    \label{fig:envmelt}
\end{figure}


As a final case study, we demonstrate the insightful application of the PDDF analysis for the characterization of the solid-liquid phase change of a nanoparticle surrounded by an interacting environment. 
Figure \ref{fig:envmelt} shows the case of a Ag$_{201}$  embedded into a strongly interacting implicit environment, modelled according to the Huerto-Cortes formalism, with $p$ and $\epsilon$ set to 1.5 and 0.04 eV/atom respectively.
The PDDF averaged over four independent melting simulations starting from a truncated octahedron and employing a heating rate of 50~K/ns and $\Delta \tau$=0.5~ns are plotted at different temperatures.
The bottom panel of Figure \ref{fig:envmelt} compares the KL(h$\mid$c) and $\Delta$ curves versus temperature.
While the caloric curve does not present any clear transition, the melting transition at $\sim$600~K is evident from the KL(h$\mid$c) analysis. Indeed, the second peak of the PDDF does not correspond to the bulk lattice distance at ``hot" temperatures. This result corroborates the strength and applicability of the melting characterization method we propose. 
We note that the observed melting temperature is much lower than in the vacuum condition case.
Calculations across a comprehensive set of metals and metal-environment parametrizations (Figures S42-S47) confirm both that i) low-symmetry structures are more likely to be observed in the solid-phase in the presence of a strongly interacting environment, since the obtained PDDF are less detailed due to the formation of defects ii) melting temperatures are consistently lower than for the case of nanoparticles in vacuum gas-phase, with stronger metal-environment interaction determining lower melting temperatures.
A deeper analysis of the effect of the environment on the melting temperature is, however, out of the aim of this paper, which focuses on identifying a general method to address the phase change at the nanoscale.

\section{Conclusion}
We present a systematic investigation of the solid-liquid phase change of several metallic nanoparticles over different sizes and shapes.
Consequently, we propose a universal signature, 
to identify the melting transition even in metallic nanoparticles in vacuo and embedded in a strongly interacting environment. 
Indeed, we show that the second peak of the pair-distance distribution function disappears in correspondence to the melting phase change.
Furthermore, we show that the relative cross-entropy of the PDDF up to the second nearest-neighbour of “cold” and “hot” configurations displays a quasi first-order transition at the melting temperature and it provides a quantitative description of the melting transition alternative to caloric curves.
To support our claim we contrast the melting temperature obtained from the caloric curve, and from the analysis of pair-distance distribution functions. Both display a perfect agreement.

Having established a clear dependence of the PDDF peaks on the system temperature paves the way towards the experimental measurement of phase changes temperatures using an alternative approach to calorimetric tools. Such analysis can lead to key development in the determination of liquid- and solid-phases for the case of systems whose heat capacity vs temperature curves result in a non-trivial interpretation, as for the case of nanoparticles in a strongly interacting environment.

\section{Acknowledgements}
LDC is supported by King's College London through the NMS Faculty Studentship Scheme. 
FB thanks the “Towards an Understanding of Catalysis on Nanoalloys” (TOUCAN) EPSRC Critical Mass Grant (EP/J010812/1) as does KR (ER/M506357/1). 
RPM and FB thank the EPSRC impact acceleration IAA award grant reference number EP/R511559/1.
PS was supported by a King's College Undergraduate Research Fellowship (KURF-2019).
All the authors are grateful to the Royal Society, RG 120207 and the IT support offered by the NMS Faculty at the King's College London. The authors acknowledge use of the research computing facility at King’s College London, Rosalind (https://rosalind.kcl.ac.uk).

\bibliography{library}
\bibliographystyle{aipnum4-1}

\clearpage

\end{document}